\documentclass[prd,aps,amsfonts,showpacs,longbibliography,notitlepage,twocolumn,groupedaddress,nofootinbib,superscriptaddress]{revtex4-1}

\usepackage{preamble}

\begin{document}
\title{Hydrodynamics as the effective field theory of strong-to-weak spontaneous symmetry breaking}

\author{Xiaoyang Huang}
\affiliation{Department of Physics and Center for Theory of Quantum Matter, University of Colorado, Boulder, CO 80309, USA}
\affiliation{Kavli Institute for Theoretical Physics, University of California, Santa Barbara, CA 93106, USA}

\author{Marvin Qi}
\affiliation{Department of Physics and Center for Theory of Quantum Matter, University of Colorado, Boulder, CO 80309, USA}

\author{Jian-Hao Zhang}
\affiliation{Department of Physics and Center for Theory of Quantum Matter, University of Colorado, Boulder, CO 80309, USA}

\author{Andrew Lucas}
\email{andrew.j.lucas@colorado.edu}
\affiliation{Department of Physics and Center for Theory of Quantum Matter, University of Colorado, Boulder, CO 80309, USA}

\begin{abstract}
Inspired by the hunt for new phases of matter in quantum mixed states, it has recently been proposed that the equivalence of microcanonical and canonical ensembles in statistical mechanics is a manifestation of strong-to-weak spontaneous symmetry breaking (SWSSB) in an underlying many-body quantum description.  Here, we build an effective field theory for SWSSB of a global U(1) symmetry; the answer exactly reproduces the Schwinger-Keldysh effective field theory of diffusion for the conserved charge.  We conclude that hydrodynamics can be understood as a theory of ``superfluidity" for the broken strong symmetry: a non-vanishing susceptibility is a measurable order parameter for SWSSB, the diffusion mode is the Goldstone boson of the spontaneously broken continuous symmetry,  and a generalization of Goldstone's Theorem implies that the diffusion mode is always long-lived.   This perspective provides a transparent physical explanation for the unusual ``reparameterization" symmetries which are a necessary ingredient of Schwinger-Keldysh effective field theories for ``normal fluids".   
\end{abstract}

\date{\today}

\maketitle

\section{Introduction} 
Hydrodynamics describes the dynamics of generic interacting many-body systems on long time and length scales.  The degrees of freedom in hydrodynamics are conventionally said to be associated to all of the conserved quantities in the system, such as charge, energy or momentum, together with Goldstone bosons for any spontaneously broken symmetries.  The past decade has seen a resurgence of work \cite{Crossley:2015evo,Glorioso:2017fpd,Haehl:2015foa,Jensen:2017kzi,Liu:2018kfw} on hydrodynamics as a formal effective field theory (EFT), built systematically from a Lagrangian made out of symmetry-invariant building blocks.  For the simplest theories of hydrodynamics, such as Fick's Law for the diffusion of a single conserved U(1) charge, this EFT reproduces well-established physics.  In more exotic theories \cite{Landry:2019iel,Gromov:2020yoc,Huang:2022ixj,Farrell:2022xnf,Glorioso:2023chm,Jain:2023nbf}, such EFT principles have played a critical role in carefully analyzing and understanding the possible hydrodynamic phenomena consistent with nontrivial spacetime symmetries and their spontaneous breaking.  

Let us review, at a high level, the way that the EFT of hydrodynamics works, for the simplest theory of a conserved U(1) charge; we follow the discussion of \cite{Crossley:2015evo}.  The concrete goal of the EFT is to calculate correlation functions of the conserved U(1) current $J^\mu = (n, \mathbf{J})$, which obeys $\partial_\mu J^\mu = 0$.  Our goal is to calculate a generating function, which takes the schematic form \begin{equation}
    Z[A_1,A_2] = \mathrm{tr}\left[ \rho \mathrm{e}^{\mathrm{i}\int J^\mu A_{\mu 1}}\mathrm{e}^{-\mathrm{i}\int J^\mu A_{\mu 2}} \right],
\end{equation}where $\rho$ represents the thermal density matrix, and we have introduced two probe fields $A_{1,2}$ because (\emph{1}) hydrodynamic correlation functions are often probed using retarded Green's functions, which are defined in terms of unequal time  commutators which cannot be calculated in a time-ordered path integral, and (\emph{2}) mixed state $\rho$ involves both a bra and a ket, which each are sourced separately.  This is the standard Schwinger-Keldysh formalism.  The key insight of the EFT is that the presence of gapless hydrodynamic modes renders this generating functional nonlocal in space and time.  To obtain a local EFT, one integrates in two scalar fields $\phi_{1,2}$: \begin{equation}
    A_{\mu \alpha} \rightarrow A_{\mu \alpha} + \partial_\mu \phi_\alpha \;\;\; (\alpha=1,2).
\end{equation}
The EFT asserts that \begin{equation}
    Z[A_1,A_2] = \int \mathrm{D}\phi_\alpha \; \mathrm{e}^{\mathrm{i}S[A_\alpha + \partial \phi_{\alpha}]},
\end{equation}
where $S$ is now a local functional that includes all allowed terms subject to various symmetries and/or constraints.  Most of these constraints are relatively well-motivated; here we will focus on the one which, in our view, is not.  The standard prescription in the literature has been to demand that \begin{equation}
    S[\phi_1,\phi_2] = S[\phi_1 + c_1 + f(x), \phi_2 + c_2 + f(x)]
\end{equation}
where $c_{1,2}$ are constants, but $f(x)$ is an arbitrary time-independent function.  The $f(x)$-shift symmetry, which we will refer to as a reparameterization symmetry, is to be imposed so long as the U(1) symmetry is not spontaneously broken -- i.e., in a normal fluid phase.  In contrast, if we study a U(1) superfluid phase, then we demand only the constant $c_{1,2}$ shift symmetries.  It is clear from the form of the EFT, which we will review later, that this prescription is the right one, but its physical motivation has appeared relatively ad hoc.

The purpose of this paper is to explain a clear physical origin for the reparameterization symmetry.  Our key insight comes from recent literature \cite{de_Groot_2022, Ma_2023, zhang2024strange, lee2024SPT, ma2024average, Zhang_2023, Chen_2024, chen2024separability, sang2023mixed, sang2024stability, ma2024double, xue2024tensor, guo2024LPDO, Lee_2023, Sala:2024ply,lessa2024strongtoweak} which uses insight from quantum information theory and (symmetry-protected) topological phases of quantum matter to classify mixed states.  A first crucial observation is that in a mixed state $\rho$ with symmetry $G$, there are two different ways to implement that symmetry.  Letting $U$ represent a unitary matrix representing one of the elements of $G$, we say that $\rho$ is strongly symmetric if \begin{equation}
    U\rho = \mathrm{e}^{\mathrm{i}\theta}\rho, \;\;\; \rho U^{-1} = \rho \mathrm{e}^{-\mathrm{i}\theta}, \label{eq:strongsymdef}
\end{equation}
and any choice of $U$ at most transforms the state $\rho$ by a phase.  In contrast, we could also have a weak symmetry: \begin{equation}
    U\rho U^{-1}= \rho,
\end{equation}
but \eqref{eq:strongsymdef} does not hold.   If we study quantum dynamics on a pure state $|\psi\rangle$, generated by symmetric Hamiltonian $H$ obeying $[H,U]=0$, then the generator of time evolution on the mixed state $\rho$ is strongly symmetric.

The hydrodynamic EFT for the conserved U(1), sketched above, describes the relaxation to a thermal state, such as
\begin{equation}
    \rho_\mu = \mathcal{Z}^{-1} \mathrm{e}^{-\mu Q}, \label{eq:gce}
\end{equation}
where $\mathcal{Z}$ is the partition function and $Q$ represents a conserved U(1) charge, i.e. the generator of the global U(1) symmetry.   It is easy to see that $\rho_\mu$ is weakly symmetric: $\mathrm{e}^{\mathrm{i}\alpha Q} \rho_\mu \mathrm{e}^{-\mathrm{i}\alpha Q} = \rho_\mu$; it is not, however, strongly symmetric: $\mathrm{e}^{\mathrm{i}\alpha Q} \rho_\mu\ne \mathrm{e}^{\mathrm{i}\theta} \rho_\mu$, since the grand canonical ensemble averages over different charge sectors.  

Statistical mechanics tells us that the hydrodynamic EFT generically applies even if we start in an initial state with known global charge: namely, the initial state does have strong symmetry.  This can be understood as the statement that physics is insensitive to the microcanonical vs. (grand) canonical ensemble in the thermodynamic limit, which has a nice interpretation in quantum many-body systems via the eigenstate thermalization hypothesis \cite{deutsch,srednicki}.   In any pure state where the diffusive hydrodynamic EFT applies, we must have $[H,U]=0$ for any $U\in \mathrm{U}(1)$, so we conclude that: (\emph{1}) the mixed state dynamics must be strongly symmetric; (\emph{2}) the state to which we locally relax (i.e. which captures local correlation functions) is not invariant under strong symmetry.  Taken together, these two facts imply that a symmetry has spontaneously broken, and we will call this phenomenon strong-to-weak spontaneous symmetry breaking (SWSSB) \cite{Sala:2024ply,lessa2024strongtoweak}.

This paper synthesizes the story of SWSSB with the EFT of hydrodynamics, focusing on the simplest case of diffusion for clarity, although our punchline should directly generalize to other hydrodynamic models.   Our key observation, which we will carefully and pedagogically motivate, is that SWSSB quantitatively justifies the reparameterization symmetry of the EFT, providing a physical justification for the one ``mysterious" aspect of the prescription of \cite{Crossley:2015evo}.  Aspects of this idea, in particular an identification of broken strong symmetry in hydrodynamics,  can be found in earlier work \cite{Akyuz:2023lsm,Ogunnaike:2023qyh,Moudgalya:2023yon}.  Our paper will start by deducing the most general effective field theory for ``superfluidity" in a theory with SWSSB, and derive the corresponding EFT, which exactly reproduces hydrodynamics \cite{Crossley:2015evo}.    Along the way, we will also propose a physical interpretation for the integrated-in $\phi$ fields as quantum channels, which may be of independent interest.

Our work also provides a concrete answer to a puzzle raised in \cite{Sala:2024ply,lessa2024strongtoweak}: if a continuous symmetry has SWSSB, is there an analogue of Goldstone's Theorem?  Our answer is affirmative: we will see that the hydrodynamic diffusion mode of the EFT is precisely the Nambu-Goldstone boson of SWSSB.  The ``gaplessness" of this mode is mandated by an appropriate generalization of Goldstone's Theorem to SWSSB. 

For the rest of the paper, we have aimed to present a pedagogical understanding of these conclusions.  In Section \ref{sec:review}, we first review the EFT of spontaneous symmetry breaking (SSB) in pure states.  Section \ref{sec:swssb} then uses these reviewed methods to define and interpret SWSSB in mixed states.  Section \ref{sec:hydro} then builds an EFT for SWSSB, which is easily seen to reproduce diffusion, along with the hydrodynamic EFT of \cite{Crossley:2015evo}, while Section \ref{sec:sf} briefly remarks on the derivation of dissipative superfluid hydrodynamics when both strong and weak symmetries spontaneously break.

\section{Review: SSB in pure states} 
\label{sec:review}
We first review the established theory of SSB in a pure state of a quantum system in $d$ spatial dimensions. 
For simplicity we will assume rotational symmetry, although it could be straightforwardly relaxed.  
Suppose that we have a Hamiltonian $H$ which is invariant under symmetry group $G$: the unitary $U(g)$ obeys $[H,U(g)]=0$ for all $g\in G$.  If $|\psi\rangle$ is an eigenstate of $H$, we say that $|\psi\rangle$ exhibits SSB if and only if \begin{equation}
    U(g)|\psi\rangle \ne \mathrm{e}^{\mathrm{i}\theta}|\psi\rangle
\end{equation}for some real $\theta$.  In words, the state $|\psi\rangle$ is not invariant under the symmetry group of the dynamics (i.e. Hamiltonian).

Of interest to us will be SSB of a continuous symmetry.  The simplest choice is then to consider $G=\mathrm{U}(1)$, with \begin{equation}
    Q = \int \mathrm{d}^dx \; n(x) \label{eq:Qint}
\end{equation} the generator of the symmetry group:  $U(\alpha) = \mathrm{e}^{\mathrm{i}\alpha Q}$ represents $G$ on the Hilbert space.  Here $Q$ denotes total charge, while $n$ denotes charge density.  If \begin{equation}
    |\psi\rangle \ne |\psi_\alpha\rangle := U(\alpha)|\psi\rangle,
\end{equation}then there should be some physical observable $A$ 
that distinguishes the two states: \begin{equation}
    \left.\frac{\mathrm{d}}{\mathrm{d}\alpha } \langle \psi | \mathrm{e}^{-\mathrm{i}\alpha Q} A \mathrm{e}^{\mathrm{i}\alpha Q}|\psi\rangle \right|_{\alpha=0} =-\mathrm{i}\langle \psi|[Q,A]|\psi\rangle \ne 0. \label{eq:ddalpha}
\end{equation}
We will take $A=\mathrm{e}^{\mathrm{i}\phi}$, where $\phi$ is a canonical conjugate of the charge density, 
\begin{equation}
     [\phi(x),n(y)] = \mathrm{i}\delta^{(d)}(x-y). \label{eq:canonical}
\end{equation}
Since U(1) is compact, $\phi \sim \phi+2\pi$, so only $A$ is well-defined -- strictly speaking, \eqref{eq:canonical} is not.  From \eqref{eq:Qint}, \eqref{eq:ddalpha} and \eqref{eq:canonical}, \begin{equation}
    \langle \psi_\alpha| \mathrm{e}^{\mathrm{i} \phi(x)} |\psi_\alpha\rangle = C\mathrm{e}^{-\mathrm{i}(\alpha-\alpha_0)}. \label{eq:alpha0}
\end{equation}
up to an integration constant $\alpha_0$.  The constant here $C>0$ when there is SSB, while if $C=0$ the state is symmetric.  

We can now deduce an EFT for these degrees of freedom.  From \eqref{eq:canonical}, $n$ and $\phi$ are canonical conjugates, so we can write a ``Hamiltonian" effective theory by choosing Lagrangian \begin{equation}
    \mathcal{L} = n\partial_t \phi - \mathcal{H}(n,\nabla n, \nabla \phi, \ldots).
\end{equation}
Due to the freedom to shift integration constant in \eqref{eq:alpha0}, $\mathcal{L}$ must be invariant under the shift symmetry \begin{equation}
    \phi \rightarrow \phi + \alpha_0, \label{eq:phishift}
\end{equation}
and this is why $\mathcal{H}$ cannot depend on $\phi$.    
The shift symmetry \eqref{eq:phishift} is the nonlinear realization of U(1) in the EFT.  

If $C>0$ in \eqref{eq:alpha0}, then (assuming translation symmetry of the states $|\psi_\alpha\rangle$) the integration constant $\alpha_0$ can only be chosen once, and must be the same for all $x$.   At lowest order in derivatives, the effective Hamiltonian is
\begin{equation} \label{eq:SFHam}
    \mathcal{H} = \epsilon(n) + \frac{K(n)}{2} (\nabla \phi)^2 + \cdots.
\end{equation}
and only depends on $\nabla \phi$.  Alternatively, we note that $\mathcal{H}$ must be invariant under the U(1) transformation \eqref{eq:phishift} in order for the charge density $n$ to be the corresponding Noether charge density. Having fixed $\mathcal{H}$, it is conventional to then integrate out $n$, which enforces \begin{equation}
    \partial_t \phi = \mu \equiv \frac{\mathrm{d}\epsilon}{\mathrm{d}n}. \label{eq:1stlaw}
\end{equation} Here $\mu$ is the chemical potential; \eqref{eq:1stlaw} is the first law of thermodynamics (at zero temperature). Presuming that we can invert the thermodynamic relation to express $n$ in terms of $\mu$, we then obtain an EFT for a superfluid phase.  Expanding around the background $\phi = \mu_0t +\tilde \phi$,  \begin{equation}
    \mathcal{L} = \frac{\chi}{2}(\partial_t \tilde \phi)^2 - \frac{K_0}{2} (\nabla \tilde\phi)^2 + \cdots. \label{eq:SFL}
\end{equation}
where we defined the positive constants $K_0=K(\mu_0)$, and $\chi^{-1} = \epsilon^{\prime\prime}(\mu_0)$ is the charge susceptibility.

In contrast, when $C=0$ and we do not have SSB, then we must set $K_0=0$.  After all, if $C=0$, then $\alpha-\alpha_0$ in \eqref{eq:alpha0} is ill-defined and can take on any value.  When we build an EFT around such a normal fluid state with $C=0$, which does not have SSB, we must impose a stronger symmetry than \eqref{eq:phishift}: the reparameterization symmetry \begin{equation}
    \phi \rightarrow \phi + f(x) \label{eq:reparameterization}
\end{equation}
for any $t$-independent function $f$.  This enforces that $\mathcal{H} = \mathcal{H}(n)$ does not depend at all on $\phi$, which is in turn a consequence of the fact that (outside of  a ``healing length") there is no energy penalty to phase fluctuations in a normal fluid.  The resulting Hamiltonian EFT is therefore trivially integrable: $\partial_t n = 0$ and $\partial_t \phi$ is independent of $t$. If the system is anomalous, there is an additional term $\mathcal{L}_{\mathrm{anom}}\sim c \p_t \phi \p_x \phi$ in $d=1$ spatial dimension, that is allowed, as it shifts by a total derivative under \eqref{eq:reparameterization}; see \appref{app:dirac}. Since we are not interested in the anomaly, we will set the coefficient $c=0$ for the remainder of the paper.

The Mermin-Wagner Theorem \cite{merminwagner} tells us that this EFT of superfluidity is, at zero temperature, only valid in $d>2$ spatial dimensions.  Using the leading order EFT, we calculate \begin{equation}
    \int \mathrm{d}t \; \langle \tilde \phi(x,t)\tilde \phi(0)\rangle \propto \int \mathrm{d}^dk \; \frac{\mathrm{e}^{\mathrm{i}k\cdot x }}{k^2} \sim |x|^{-(d-2)},
\end{equation}
with $\log |x|$ in $d=2$, which implies that if $d\le 2$, correlation functions of $\phi$ grow with distance and hence $\phi$ cannot be long-range ordered at different points in space.  While this is commonly taken to demonstrate the absence of SSB, we note that there do exist EFTs of SSB phases without long-range order \cite{Glorioso:2023chm,Jain:2023nbf}, and so this ``lore" is incorrect in general.

It is not accidental that \eqref{eq:SFL} describes a gapless degree of freedom -- the Nambu-Goldstone boson, with dispersion relation \begin{equation}
    \omega \approx \pm \sqrt{\frac{K_0}{\chi}} |k|. \label{eq:ngb}
\end{equation}  The existence of this mode is guaranteed by Goldstone's Theorem, which can be understood as follows.  Using \eqref{eq:canonical}, \begin{align}
    \int \mathrm{d}^dx \; G^{\mathrm{R}}_{nA}(x,t)&=\mathrm{i}\mathrm{\Theta}(t) \int \mathrm{d}^dx \left\langle [n(x,t),A(0,0)] \right\rangle \notag \\
    &= \mathrm{i}\mathrm{\Theta}(t) \langle [Q,A(0)]\rangle = \mathrm{i}\mathrm{\Theta}(t) \langle A\rangle . \label{eq:goldstone0}
\end{align} 
Here $A=\mathrm{e}^{\mathrm{i}\phi}$, and expectation value is taken in the state $|\psi_{\alpha=0}\rangle$, for example, with real $\langle A\rangle$.   This implies that \begin{equation}\label{eq:Gr peak}
    G^{\mathrm{R}}_{nA}(k=0,\omega) = \left(\mathrm{i}\pi \delta(\omega) - \frac{1}{\omega}\right)\langle A\rangle.
\end{equation}
The singular structure in the spectral weight of this Green's function at $\omega=0$ necessitates the existence of a gapless mode \eqref{eq:ngb} in the spectrum independently of microscopic details. 

\section{SWSSB from mixed states}
We now turn to the problem of SWSSB in mixed states.  In this section, we first describe a short argument for SWSSB from a single-copy Hilbert space. 

Suppose our system undergoes dynamics (e.g. time-evolution by a Lindbladian) which is invariant under a strong continuous symmetry $G$. In the case of interest where $G=\mathrm{U}(1)$, this corresponds to dynamics where the particle number is separately conserved in both the system and environment. We will restrict to $G = \mathrm{U}(1)$ in the remainder of the paper. If $\rho$ is a steady state of the dynamics, we say that $\rho$ exhibits SWSSB if and only if the weak symmetry is preserved,
\begin{equation}
    U(\alpha) \rho U(\alpha)^{-1} = \rho,
\end{equation}
and the strong symmetry is broken,
\begin{equation}
    U(\alpha) \rho \neq \mathrm{e}^{\mathrm{i} \theta} \rho,
\end{equation}
with $U(\alpha) = \mathrm{e}^{\mathrm{i} \alpha Q}$ and $Q$ defined as in \eqref{eq:Qint}. 

For any steady state density matrix $\rho$, the family of states 
\begin{equation}
    \rho_\alpha = \frac{1}{\mathrm{Tr}(\mathrm{e}^{\alpha Q} \rho)} e^{\alpha Q} \rho 
\end{equation}
are also steady state density matrices. If $\rho_\alpha \neq \rho$, then there should be a local operator $A$ whose expectation value distinguishes the two states:
\begin{equation}\label{eq:dAdalpha}
\begin{aligned}
    \left.\frac{\mathrm{d}}{\mathrm{d}\alpha} \langle A \rangle_\alpha \right|_{\alpha=0} 
    &= \left.\frac{\mathrm{d}}{\mathrm{d}\alpha} \frac{\mathrm{Tr} (A \rho_\alpha)}{\mathrm{Tr}\rho_\alpha} \right|_{\alpha=0} \\
    &= \mathrm{Tr}(A Q \rho) - \mathrm{Tr}{Q \rho} \mathrm{Tr}{A \rho} \\
    &= \langle A Q \rangle - \langle A \rangle \langle Q \rangle \\
    &= \frac{1}{2} \langle \{A, Q \} \rangle - \langle A \rangle \langle Q \rangle
\end{aligned}
\end{equation}
where the last expression follows because $[Q, \rho] = 0$, i.e. the weak symmetry is unbroken. If we take $A = n$ to be the local charge density, we see that this quantity can be understood as the charge susceptibility $\chi$. The charge susceptibility can therefore be identified with the order parameter of SWSSB: a nonvanishing charge susceptibility implies that $\rho_\alpha \neq \rho$, indicating that the strong symmetry is spontaneously broken.

We now deduce the analogue of Goldstone's Theorem for SWSSB. Using \eqref{eq:Qint} and the fact that $Q = Q(t)$ is conserved, let us rewrite the susceptibility as 
\begin{equation}
\begin{aligned}
    \chi &= \frac{1}{2}  \langle \{n(0), Q(t) \} \rangle - \langle n \rangle \langle Q(t) \rangle \\
    &= \int \mathrm{d}^d x \;  \frac{1}{2} \langle \{n(0), n(x,t) \} \rangle - \langle n \rangle^2 \\
    &= \int \mathrm{d}^d x \; \frac{1}{2} \langle \{ n(0) - \bar{n}, n(x,t) - \bar{n} \} \rangle \\
    &= \int \mathrm{d}^d x \; G^{\mathrm{S}}_{nn}(x,t)
\end{aligned}
\end{equation}
where $G^{\mathrm{S}}_{nn}$ is the symmetric Green's function. Fourier transforming in time, we obtain 
\begin{equation}
    2 \pi \chi_{nn} \delta(\omega) = G^{\mathrm{S}}_{nn} (k = 0, \omega).
\end{equation}
We see that a nonvanishing susceptibility implies a nontrivial pole structure for the symmetric Green's function $G^S_{nn}$. In Section \ref{sec:hydro}, we will demonstrate how this structure is reflected in the effective field theory. 

\section{SWSSB in a ``doubled" Hilbert space}\label{sec:swssb}
To make contact with the effective field theory and identify the Goldstone mode, it will be useful to pass from mixed state density matrices to pure states on the ``doubled Hilbert space": \begin{equation}
    |a\rangle \langle b| \rightarrow |ab),
\end{equation}
where $|ab)\equiv |a\rangle \otimes |b\rangle^*$ with $|b\rangle^*$ being the complex conjugation of the state $|b\rangle$.
The parentheses bra-ket notation emphasizes that we are in the doubled Hilbert space.  The inner product on the doubled Hilbert space is \begin{equation}
    (\rho_1|\rho_2) = \mathrm{tr}\left(\rho_1^\dagger \rho_2\right).
\end{equation} We will also define left (L) and right (R) operators that act as follows: \begin{subequations}
    \begin{align}
        A_{\mathrm{L}}|ab) &:= A|a\rangle\langle b|, \\
        B_{\mathrm{R}}|ab) &:= |a\rangle\langle b|B^{\mathsf{T}},
    \end{align}
\end{subequations}
where we used the map $|a\rangle \otimes B|b\rangle^* \to |a\rangle \langle b| B^{\mathsf{T}}$.
Physical dynamics in the doubled Hilbert space, in continuous time, is generated by Lindbladian $\mathcal{H}$ (this is not the standard letter used, but will avoid clashes in Section \ref{sec:hydro}), which is a completely-positive trace-preserving (CPTP) map.  CP does not provide a linear constraint on $\mathcal{H}$, but TP does: letting $|I)$ denote the identity operator, \begin{equation}
    \mathcal{H}^\dagger|I) = 0. \label{eq:tracepreserving}
\end{equation}
From the doubled Hilbert space perspective, we will need to build $\mathcal{H}$ consistent with \eqref{eq:tracepreserving}, which we will see generally leads to a non-Hermitian $\mathcal{H}$.  Physical correlation functions are measured with $(I|\cdots |\rho)$, which has many consequences that we will later unpack. 

With this notation in mind, let us consider dynamics that protects the mixed state of a system with a conserved U(1) symmetry in the grand canonical ensemble \eqref{eq:gce}. For concreteness in the discussion that follows, it helps to consider the explicit setting of a lattice model with quantum rotor degrees of freedom $n_i$, $\phi_i$ on lattice sites, with $[\phi_i,n_j] = \mathrm{i}\delta_{ij}$.  As before, $\phi_i \sim \phi_i + 2\pi$, so $n_i\in\mathbb{Z}$.  A useful basis for the Hilbert space on each site is simply $|n\rangle$, and $\mathrm{e}^{\mathrm{i}p\phi}|n\rangle = |n+p\rangle$, with $\mathrm{e}^{\mathrm{i}p\phi}$ only defined for integer $p$.   The generator of the conserved U(1) on the single-copy Hilbert space is
\begin{equation}
    Q = \sum_i n_i.
\end{equation}
Suppose that the dynamics drives the quantum system to the steady-state
\begin{equation}
    \rho = \frac{1}{\mathcal{Z}} \mathrm{e}^{-\Phi} = \frac{1}{\mathcal{Z}} \exp\left[-\sum_i \varphi(n_i) \right] \label{eq:definerho}
\end{equation}
with $\mathcal{Z}=\mathrm{tr}( \mathrm{e}^{-\Phi})$ and $\varphi$ a function which tends to $+\infty$ as $n\rightarrow \pm \infty$. 
In the doubled Hilbert space picture, 
\begin{equation}
    \rho = \frac{1}{\mathcal{Z}}\bigotimes_i \sum_{n\in\mathbb{Z}} \mathrm{e}^{-\varphi(n)} |nn)_i.
\end{equation} We will find it useful to define a ``chemical potential" operator: 
\begin{align}
    \mu_i &= \mathrm{e}^{-\mathrm{i}\phi}\varphi(n_i) \mathrm{e}^{\mathrm{i}\phi} - \varphi(n_i) \notag \\
    &= \varphi(n_i+1)-\varphi(n_i) \approx \frac{\partial \Phi}{\partial n_i}. \label{eq:chemicalpotential}
\end{align}
The last approximation holds if $\varphi$ is slowly-varying and $n_i$ is relatively large; this semiclassical limit will connect very concretely to our EFT later.

We consider strongly-symmetric dynamics generated by a Lindbladian $\mathcal{H}$, which has a strong $\mathrm{U}(1)$ symmetry, meaning that in the doubled Hilbert space, 
\begin{equation}
    [\mathcal{H}, Q_\mathrm{L}] = [\mathcal{H}, Q_\mathrm{R}] = 0.
\end{equation}
Notice that the steady state $|\rho)$ is \emph{not} invariant under the strong symmetry:  \begin{equation}
    \mathrm{e}^{\alpha Q_{\mathrm{L}}} |\rho) = \mathrm{e}^{\alpha Q_{\mathrm{R}}} |\rho) \propto |\rho_\alpha), \label{eq:alphaQL}
\end{equation}
where $\rho_\alpha$ has $\varphi \rightarrow \varphi - \alpha n_i$. 
This equation is not an equality because the density matrix does not stay normalized.
On the other hand, $|\rho)$ \emph{is} invariant under the weak U(1) symmetry generated by $Q_{\mathrm{L}}-Q_{\mathrm{R}}$: \begin{equation}
    (Q_{\mathrm{L}}-Q_{\mathrm{R}}) |\rho) = 0.
\end{equation}
Therefore, we have SWSSB of $\mathrm{U}(1)_{\mathrm{L}}\times \mathrm{U}(1)_{\mathrm{R}}$ being broken to the weak $\mathrm{U}(1)_{\mathrm{r}}$ symmetry, where we have defined a convenient basis of r (weak) and a (strong) operators \begin{subequations}\label{eq:arQ}
    \begin{align}
        Q_{\mathrm{a}} &:= \frac{Q_{\mathrm{L}}+Q_{\mathrm{R}}}{2}, \\
        Q_{\mathrm{r}} &:= Q_{\mathrm{L}}-Q_{\mathrm{R}}.
    \end{align}
\end{subequations}
The labels a/r refer to strong/weak symmetries for later comparison with the KMS-invariant EFT \eqref{eq:SK}.  

Strictly speaking, to justify the existence of SSB in the thermodynamic limit, we should replace the grand canonical $\rho$ from \eqref{eq:definerho} with \begin{equation}
    \tilde \rho  \propto  \delta_{\sum n_i, N} \rho. \label{eq:tilderho}
\end{equation}
Notice that $\rho$ and $\delta_{\sum n_i, N}$ commute, and that $\mathrm{e}^{\mathrm{i} \alpha Q_{\mathrm{L}}}|\tilde\rho) = \mathrm{e}^{\mathrm{i} \alpha N} |\tilde\rho)$, meaning that $|\tilde\rho)$ is strongly symmetric as it shifts only by a phase. We will later carefully explain why $|\tilde \rho)$ indeed exhibits SWSSB, but intuitively, we notice that statistical mechanics suggests that there is no physical way to tell $|\rho)$ and $|\tilde\rho)$ apart when studying local observables.  For this reason, it makes sense to work with the original state $|\rho)$ which is not strongly symmetric.

Ultimately our goal is to build a Hamiltonian EFT of this SWSSB ``phase".  Following Section \ref{sec:review}, we will build an EFT in terms of the local densities $n_{\mathrm{a}}$ and $n_{\mathrm{r}}$ along with their canonical conjugates $\phi_{\mathrm{a}}$ and $\phi_{\mathrm{r}}$.  More precisely, since $\phi_{\mathrm{a}}$ and $\phi_{\mathrm{r}}$ are not well-defined, we will work with the operators \begin{subequations} \label{eq:phiaphirdef}
    \begin{align}
        \mathrm{e}^{\mathrm{i}\phi_{\mathrm{a}}} &:= \left(\mathrm{e}^{\mathrm{i}\phi}\right)_{\mathrm{L}}\left(\mathrm{e}^{\mathrm{i}\phi}\right)_{\mathrm{R}} , \\         \mathrm{e}^{-\mathrm{i}\phi_{\mathrm{a}}} &:= \left(\mathrm{e}^{-\mathrm{i}\phi}\right)_{\mathrm{L}}\left(\mathrm{e}^{-\mathrm{i}\phi}\right)_{\mathrm{R}} , \\        
        \mathrm{e}^{2\mathrm{i}\phi_{\mathrm{r}}} &:= \left(\mathrm{e}^{\mathrm{i}\phi}\right)_{\mathrm{L}}\left(\mathrm{e}^{-\mathrm{i}\phi}\right)_{\mathrm{R}} , \\         \mathrm{e}^{-2\mathrm{i}\phi_{\mathrm{r}}} &:= \left(\mathrm{e}^{-\mathrm{i}\phi}\right)_{\mathrm{L}}\left(\mathrm{e}^{\mathrm{i}\phi}\right)_{\mathrm{R}}.
    \end{align}
\end{subequations}
Notice that $\mathrm{e}^{\pm \mathrm{i}\phi_{\mathrm{a}}}|nn) = |(n\pm 1)(n\pm 1))$ and that these definitions are for single site operators.  Normalizations are chosen so that a/r variables obey the same rotor algebra as before.  We also have the useful identity \begin{equation}
    (I| \mathrm{e}^{\pm \mathrm{i}\phi_{\mathrm{a}}} = (I|, \label{eq:phiachannel}
\end{equation} 
which is why we have chosen these definitions for $\phi_{\mathrm{a,r}}$ over others that would have obeyed the same algebra: \begin{subequations}
    \begin{align}
        [n_{\mathrm{a}},\mathrm{e}^{\mathrm{i}\phi_{\mathrm{a}}}] &= \mathrm{e}^{\mathrm{i}\phi_{\mathrm{a}}}, \\
        [n_{\mathrm{r}},\mathrm{e}^{2\mathrm{i}\phi_{\mathrm{r}}}] &= 2\mathrm{e}^{2\mathrm{i}\phi_{\mathrm{r}}}, \\
        [n_{\mathrm{a}},\mathrm{e}^{2\mathrm{i}\phi_{\mathrm{r}}}] =  [n_{\mathrm{r}},\mathrm{e}^{\mathrm{i}\phi_{\mathrm{a}}}]  &= 0.
    \end{align}
\end{subequations}
Crucially, \eqref{eq:phiachannel} implies that $\mathrm{e}^{\pm \mathrm{i}\phi_{\mathrm{a}}}$ is a quantum channel, not an operator acting on the single-copy Hilbert space.  Lastly, explicit calculation confirms that (for a single site)
\begin{subequations}
    \begin{align}
        (I|n_{\mathrm{a}}|\rho) &= \mathrm{tr}(n \rho), \\
        (I| \mathrm{e}^{\mathrm{i}\phi_{\mathrm{a}}}|\rho) &= (I|\rho) = 1, \label{eq:navev} \\
        (I| n_{i\mathrm{r}} |\rho) &= 0. \label{eq:nr0}
    \end{align}
\end{subequations}
$\phi_{\mathrm{r}}$ does not have a well-defined value in our steady state, as discussed around \eqref{eq:reparameterization}.

It will be important to understand the time-reversal transformation in our system.    Following \cite{Guo:2024ycr}, notice that given any physical observable, we may always write \begin{equation}\label{eq:kms}
    (I|\mathcal{A}|\rho)^* = (\rho|\mathcal{A}^\dagger|I) = (I|\mathcal{S}\mathcal{A}^\dagger \mathcal{S}^{-1}|\rho)
\end{equation}
where the first equality comes from the identity $\tr(\mathcal{A} \rho)^* = \tr(\rho^\dagger \mathcal{A}^\dagger)$,  and in the second equality we introduced
\begin{equation}
    \mathcal{S}|A):= |\sqrt{\rho}A\sqrt{\rho}).
\end{equation}
Note the transformation in \eqref{eq:kms} is sometimes called modular conjugation.  
Define the time-reversed superoperator \begin{equation}
    \tilde{\mathcal{A}} := \mathcal{S}\mathcal{A}^\dagger \mathcal{S}^{-1}.
\end{equation}
Notice that this generalized time-reversal squares to the identity, so it is a sensible candidate for time-reversal.  We can explicitly calculate that $(A_{\mathrm{L}})^\dagger = (A^\dagger)_{\mathrm{L}}$ and $(A_{\mathrm{R}})^\dagger = (A^\dagger)_{\mathrm{R}}$, and $\mathcal{S}(\mathrm{e}^{\pm \mathrm{i}\phi})_{\mathrm{L,R}} \mathcal{S}^{-1} \approx \mathrm{e}^{\mp \mu/2} (\mathrm{e}^{\pm \mathrm{i}\phi})_{\mathrm{L,R}}$, with the latter approximation reasonable so long as $\mu$ is a slowly varying function. Hence, \begin{subequations}\label{eq:TRS}
    \begin{align}
        \tilde n_{i\mathrm{a}} &=  n_{i\mathrm{a}}, \\
        \tilde n_{i\mathrm{r}} &=  n_{i\mathrm{r}}, \\
        \mathrm{e}^{\pm\mathrm{i}\tilde \phi_{i\mathrm{a}}} &= \mathrm{e}^{\pm\mathrm{i}(-\phi_{i\mathrm{a}} -\mathrm{i}\mu_i)}, \\
        \mathrm{e}^{\pm\mathrm{i}\tilde \phi_{i\mathrm{r}}} &= \mathrm{e}^{\pm\mathrm{i}(-\phi_{i\mathrm{r}})}.
    \end{align}
\end{subequations}

Now, let us explain why there is no Mermin-Wagner Theorem \cite{merminwagner} for SWSSB.  
Roughly speaking, from the eigenstate thermalization hypothesis and statistical mechanics, insofar as local correlation functions are concerned, the density matrix in our mixed steady state $\rho$ is well-approximated by a grand canonical ensemble \eqref{eq:gce}, which exhibits SWSSB.  Since \eqref{eq:gce} is a tensor product, there is no long-range order accessible in any local correlators, without using multiple copies of the density matrix.  (This renders the probes of SSB unmeasurable in current experiments.)

Let us now be a little more careful.  To see the emergence of a ``long-range order" which can be associated to SWSSB, we should start with a strongly symmetric state, and a useful choice will be \eqref{eq:tilderho}. In the thermodynamic limit, if $N = \langle n\rangle V$ (where $V$ is the number of lattice sites) is chosen such that $\langle n\rangle$ is the expectation value in the grand canonical ensemble $\rho$, defined in \eqref{eq:definerho}. To see the presence of long-range order, we must look at more complicated correlation functions involving multiple copies of a density matrix, such as the Wightman's or R\'enyi-1 correlator \cite{Wightman, weinstein2024}
\begin{align}
    F_{xy} := \mathrm{tr}&\left( \sqrt{\tilde \rho} \left(\mathrm{e}^{\mathrm{i}p\phi_{x}}\mathrm{e}^{-\mathrm{i}p\phi_{y}}\right) \sqrt{\tilde \rho} \left(\mathrm{e}^{\mathrm{i}p\phi_{x}}\mathrm{e}^{-\mathrm{i}p\phi_{y}}\right)^\dagger \right).  
    \label{eq:fidelity}
\end{align}
for some integer $p\ne 0$. Let us now argue that \begin{equation}
   F_{xy} = \delta_{xy} + a(1-\delta_{xy}), \label{eq:fidelityform}
\end{equation}
where
\begin{equation}
    a = \left(\dfrac{\displaystyle \sum_{n\in\mathbb{Z}} \mathrm{e}^{-[\varphi(n)+\varphi(n+p)]/2}}{\displaystyle \sum_{n\in\mathbb{Z}} \mathrm{e}^{-\varphi(n)}} \right)^2.
\end{equation} 
From the form of $F_{xy}$, and because $\delta_{\sum n, N}$ commutes with both $\sqrt{\rho}$ and the product $\mathrm{e}^{\mathrm{i}\phi_{x}}\mathrm{e}^{-\mathrm{i}\phi_{y}}$, we may write \begin{align}
    F_{xy} &= \mathrm{tr} \left( \delta_{\sum n, N} \sqrt{\rho}\left(\mathrm{e}^{\mathrm{i}p\phi_{x}}\mathrm{e}^{-\mathrm{i}p\phi_{y}}\right) \sqrt{ \rho} \left(\mathrm{e}^{\mathrm{i}p\phi_{x}}\mathrm{e}^{-\mathrm{i}p\phi_{y}}\right)^\dagger \right) \notag \\
    &= \int\limits_0^{2\pi}\frac{\mathrm{d}\theta}{2\pi} \mathrm{tr} \left( \mathrm{e}^{\mathrm{i}\theta (\sum n- N)}\sqrt{ \rho} \left(\mathrm{e}^{\mathrm{i}p\phi_{x}}\mathrm{e}^{-\mathrm{i}p\phi_{y}}\right)\sqrt{ \rho} \left(\mathrm{e}^{\mathrm{i}p\phi_{x}}\mathrm{e}^{-\mathrm{i}p\phi_{y}}\right)^\dagger \right) \notag \\
    &= \int\limits_0^{2\pi}\frac{\mathrm{d}\theta}{2\pi} \mathcal{Y}(\theta)^{V-2} |\tilde{\mathcal{Y}}(\theta)|^2  \label{eq:fidelityintegral}
\end{align}
where \begin{subequations}
    \begin{align}
    \mathcal{Y}(\theta) &= \dfrac{\displaystyle \sum_{n\in\mathbb{Z}} \mathrm{e}^{\mathrm{i}\theta (n-\langle n\rangle)-\varphi(n)}}{\displaystyle \sum_{n\in\mathbb{Z}} \mathrm{e}^{-\varphi(n)}}, \\
        \tilde{\mathcal{Y}}(\theta) &= \dfrac{\displaystyle \sum_{n\in\mathbb{Z}} \mathrm{e}^{\mathrm{i}\theta (n-\langle n\rangle)-[\varphi(n)+\varphi(n+p)]/2}}{\displaystyle \sum_{n\in\mathbb{Z}} \mathrm{e}^{-\varphi(n)}}.
    \end{align}
\end{subequations}
In the thermodynamic limit $V\rightarrow \infty$, the integral in \eqref{eq:fidelityintegral} can be done asymptotically by saddle point which will collapse onto $\theta\approx 0$.  This leads to \eqref{eq:fidelityform}.  In contrast, the ``disconnected" correlator \begin{equation}
    \mathrm{tr}\left(\sqrt{\tilde\rho} \mathrm{e}^{\mathrm{i}p\phi_x} \sqrt{\tilde\rho}\mathrm{e}^{-\mathrm{i}p\phi_x}\right) = 0
\end{equation}
since $\mathrm{e}^{\mathrm{i}p\phi_x} \sqrt{\tilde\rho}\mathrm{e}^{-\mathrm{i}p\phi_x}$ lies in the charge sector $N+1$ which is orthogonal to $N$.  Hence we have long-range order in $F_{xy}$.   The result does not depend on spatial dimension.  Since $\tilde \rho$ is not the ground state of a spatially-local Hermitian Hamiltonian, the conventional Mermin-Wagner Theorem has no bearing on $F_{xy}$.  Since $\rho$ is a diagonal density matrix from \eqref{eq:definerho} and remains diagonal in the same basis if we conjugate $\rho$ by the operator $\mathrm{e}^{\mathrm{i}p\phi_{x}}\mathrm{e}^{-\mathrm{i}p\phi_{y}}$, \eqref{eq:fidelity} is exactly the fidelity correlator defined in \cite{lessa2024strongtoweak} to be the universal order parameter of SWSSB: 
\begin{align}
F & \left(\rho, \left(\mathrm{e}^{\mathrm{i}p\phi_{x}}\mathrm{e}^{-\mathrm{i}p\phi_{y}}\right){\rho} \left(\mathrm{e}^{\mathrm{i}p\phi_{x}}\mathrm{e}^{-\mathrm{i}p\phi_{y}}\right)^\dagger\right)\nonumber\\
&=\mathrm{tr}\left(\sqrt{\sqrt{\rho} \left(\mathrm{e}^{\mathrm{i}p\phi_{x}}\mathrm{e}^{-\mathrm{i}p\phi_{y}}\right){\rho} \left(\mathrm{e}^{\mathrm{i}p\phi_{x}}\mathrm{e}^{-\mathrm{i}p\phi_{y}}\right)^\dagger\sqrt{\rho}}\right).
\end{align}

In general, $a=0$ only when $\Phi$ is concentrated on a single value of $n$, which describes an insulating state without SWSSB; all other states have SWSSB.
As this correlation function is not measurable in an actual many-body system, and the existence of long-range order is a property of the state $\rho$, which is a physical steady state in any spatial dimension, we again see that the Mermin-Wagner Theorem is not related to SWSSB. 

Although the fidelity correlator that detected SWSSB is not measurable in experiment, the non-vanishing of $F_{xy}$ is detectable via a very simple thermodynamic quantity: the compressibility $\chi = \partial \langle n\rangle/\partial \mu$, where $\mu$ is the thermodynamic chemical potential.  Imagine shifting $\mu \rightarrow \mu + \delta \mu$, which would adjust \begin{equation}
    \varphi(n) \rightarrow \varphi(n) + n \cdot \delta \mu. \label{eq:nshiftmu}
\end{equation}
Then \begin{equation}
    \chi = \frac{\partial \langle n\rangle}{\partial \delta \mu} = \frac{\partial}{\partial \delta \mu}\dfrac{\displaystyle \sum_{n\in\mathbb{Z}} n\mathrm{e}^{-\varphi(n)-\delta \mu n}}{\displaystyle \sum_{n\in\mathbb{Z}} \mathrm{e}^{-\varphi(n) - \delta \mu n}} = \left\langle n^2\right\rangle - \langle n\rangle^2,
\end{equation}
with averages taken in the ensemble $\rho$.  $\chi > 0$ if and only if $\varphi(n) < \infty$ on more than one value of $n$, which occurs if and only if $F_{xy}\ne 0$ for some $p$.  

If we consider the toy model \begin{equation}
    \varphi(n) \approx \frac{(n-\langle n\rangle)^2}{2\chi}
\end{equation}
where $\chi \ll 1$, such that the discretization in $n$ is not important, then we notice that (on a single lattice site) \begin{equation}
    \mathrm{e}^{\mathrm{i}\phi_{\mathrm{a}}}|\rho_{\langle n\rangle}) = |\rho_{\langle n\rangle + 1}) \propto \mathrm{e}^{n_{\mathrm{a}}/\chi}|\rho_{\langle n\rangle}). \label{eq:phiaapprox}
\end{equation}
shifts the steady state to one with a higher value of $\langle n\rangle$, which from \eqref{eq:nshiftmu} can be interpreted as adjusting the chemical potential.  This implies that in a state with SWSSB, we can mirror \eqref{eq:ddalpha} by studying \begin{equation}
    1 = (I| [n_{\mathrm{a}},\mathrm{e}^{\mathrm{i}\phi_{\mathrm{a}}}]|\rho) = \mathrm{tr}(n \rho_{\langle n\rangle +1}) - \mathrm{tr}(n\rho_{\langle n\rangle}). \label{eq:singlesite}
\end{equation}
The latter equality above only holds for states with SWSSB, while the former equality holds in all states. To return to the many-body problem, $|\rho)$ in \eqref{eq:singlesite} should be understood as the reduced density matrix on a single site.  Still, the former equality will underlie our generalization of Goldstone's Theorem to SWSSB in Section \ref{sec:hydro}, whenever $\chi>0$.

\section{Effective field theory}\label{sec:hydro}
At long last, we are ready to put together the EFT of SWSSB, following Section \ref{sec:review} while being careful about the subtleties discussed in Section \ref{sec:swssb}.  The fields in the Hamiltonian EFT are $n_{\mathrm{a}}, n_{\mathrm{r}}, \phi_{\mathrm{a}}, \phi_{\mathrm{r}}$, with Lagrangian \begin{equation}
    \mathcal{L} = n_{\mathrm{a}} \partial_t \phi_{\mathrm{a}} + n_{\mathrm{r}} \partial_t \phi_{\mathrm{r}} - \mathcal{H}(n_{\mathrm{a,r}},  \phi_{\mathrm{a,r}}, \nabla n_{\mathrm{a,r}}  \ldots). \label{eq:dissL}
\end{equation}
So far, this is not the standard Schwinger-Keldysh (SK) formalism \cite{Crossley:2015evo}, although there is a transparent way to derive the SK EFT which we  will come back to compare to soon.
There are a number of constraints on $\mathcal{H}$, which we now enumerate.

(\emph{1}) For simplicity, we assume that the EFT is time-reversal symmetric. Time-reversal acts on the fields as in \eqref{eq:TRS}: $n_{\mathrm{a,r}}$ are even, while $\phi_{\mathrm{r}} \rightarrow -\phi_{\mathrm{r}}$ and \begin{equation}
    \phi_{\mathrm{a}} \rightarrow -\phi_{\mathrm{a}}- \mathrm{i}\mu(n_{\mathrm{a}}). \label{eq:phiaTRS}
\end{equation} If $\mathcal{L}$ is invariant up to a total derivative, we must have \begin{equation}
    \mathcal{H}(n_{\mathrm{a}}, n_{\mathrm{r}},\phi_{\mathrm{a}}, \phi_{\mathrm{r}}) = \mathcal{H}(n_{\mathrm{a}}, n_{\mathrm{r}},-\phi_{\mathrm{a}} - \mathrm{i}\mu(n_{\mathrm{a}}), -\phi_{\mathrm{r}})
\end{equation}
where $\mu(n)$ is the chemical potential \eqref{eq:chemicalpotential}.

(\emph{2}) Generalizing the discussion below \eqref{eq:SFHam}, we must have a shift symmetry associated to $\mathrm{U}(1)_{\mathrm{a}}$ and $\mathrm{U}(1)_{\mathrm{r}}$; after all, neither symmetry is explicitly broken.  Thus, $\mathcal{H}$ can only depend on $\nabla \phi_{\mathrm{a,r}}$.  But if we are studying dynamics in the grand canonical ensemble $|\rho)$, $\mathrm{U}(1)_{\mathrm{a}}$ is SSB while $\mathrm{U}(1)_{\mathrm{r}}$ is not.  Following \eqref{eq:reparameterization}, we then expect to have a reparameterization symmetry for $\phi_{\mathrm{r}}$.  Hence, \begin{equation} \label{eq:reparameterization2}
    \mathcal{H}(n_{\mathrm{a}}, n_{\mathrm{r}}, \phi_{\mathrm{a}}, \phi_{\mathrm{r}}) = \mathcal{H}(n_{\mathrm{a}}, n_{\mathrm{r}}, \phi_{\mathrm{a}} + \alpha, \phi_{\mathrm{r}} + f(x))
\end{equation}
for constant $\alpha$ and function $f$.

(\emph{3}) Suppose that \begin{equation}
    \frac{\mathrm{d}}{\mathrm{d}t}|\rho) = -\mathrm{i}\mathcal{H}|\rho);
\end{equation}
as we explained in Section \ref{sec:swssb}, $\mathcal{H}$ need not be Hermitian.  In the EFT, therefore, $\mathcal{H}$ will not be real-valued.  Convergence of a path integral over $\exp[\mathrm{i}\int \mathrm{d}^dx\mathrm{d}t \; \mathcal{L}]$ evidently requires \begin{equation}
    \mathrm{Im}(\mathcal{H}) \le 0. \label{eq:positivity}
\end{equation}

 (\emph{4}) Lastly, we must account for the constraint \eqref{eq:tracepreserving}.  To see what this does, let us consider a special case where \begin{equation}
     \mathcal{H} = H(n_{\mathrm{L}}) - H(n_{\mathrm{R}}), \label{eq:HLHR}
 \end{equation}
 corresponding to purely unitary quantum dynamics generated by a density-dependent Hamiltonian $H$, appropriate to a normal fluid phase.  Notice that $\mathcal{H}^\dagger|I)=0$, which follows from $\mathcal{H}|\rho) = |[H,\rho])$.  If we had chosen $\mathcal{H}$ to depend only on $n_{\mathrm{a}}$, then $\mathcal{H}|\rho)$ would be a series of anticommutators, which would not be guaranteed to have vanishing trace.  Meanwhile, no $\phi_{\mathrm{r}}$-only terms are allowed since in \eqref{eq:phiaphirdef}, $\mathrm{e}^{\pm\mathrm{i}\phi_{\mathrm{r}}}$ changes the relative charge in the L/R Hilbert spaces, meaning that there is a vanishing inner product between $|\rho)$ and $(I|$ under dynamics.
 Therefore, we demand
\begin{equation}
     \mathcal{H}(n_{\mathrm{a}},0,0,\phi_{\mathrm{r}}) = 0, \label{eq:Hna0}
 \end{equation}
 such that every term in $\mathcal{H}$ needs at least one power of either $\nabla \phi_{\mathrm{a}}$ or $n_{\mathrm{r}}$.

 We have now enumerated the symmetries and constraints on the EFT of SWSSB.  The most general expression at leading order in fields with vanishing expectation value in equilibrium ($\nabla \phi_{\mathrm{a}}$, $n_{\mathrm{r}}$) is 
 \begin{equation}
         \mathcal{H} =   \hat\mu(n_{\mathrm{a}})n_{\mathrm{r}} - \mathrm{i}\sigma(n_{\mathrm{a}}) \nabla \phi_{\mathrm{a}} \cdot \nabla (\phi_{\mathrm{a}}+\mathrm{i}\mu(n_{\mathrm{a}})) + \cdots. \label{eq:dissH}
 \end{equation}
 where $\sigma\geq 0$ due to \eqref{eq:positivity}.  We emphasize that the coefficient $\hat\mu$ above does not formally need to equal $\mu$ from \eqref{eq:chemicalpotential}, at this point.
Combining \eqref{eq:dissL} and \eqref{eq:dissH}, we notice that $n_{\mathrm{r}}$ is a Lagrange multiplier that fixes 
\begin{equation}\label{eq:dtphir mu}
    \partial_t\phi_{\mathrm{r}} = \hat\mu(n_{\mathrm{a}}) + \cdots,
\end{equation}
where $\cdots$ denote irrelevant corrections.  Hence, we can integrate out two of our four fields.  One choice is to integrate out the r-fields, which leads to the Martin-Siggia-Rose (MSR) \cite{MSR} Lagrangian: at leading order, \begin{equation}
    \mathcal{L} = n_{\mathrm{a}}\partial_t \phi_{\mathrm{a}} + \mathrm{i}\sigma(n_{\mathrm{a}}) \nabla \phi_{\mathrm{a}}\cdot \nabla (\phi_{\mathrm{a}}+\mathrm{i}\mu(n_{\mathrm{a}})) \label{eq:MSR}
\end{equation}
Varying with respect to $\phi_{\mathrm{a}}$, we obtain the diffusion equation for the conserved charge: \begin{equation}\label{eq:diffusion}
    \partial_t n_{\mathrm{a}} = \nabla \cdot (D\nabla n_{\mathrm{a}}) + \cdots.
\end{equation}
with $D = \sigma/\chi$ the diffusion constant, and the thermodynamic stability constraint that susceptibility $\chi = \partial n_{\mathrm{a}} /\partial \mu \ge 0$.
The $\cdots$ include $\phi_{\mathrm{a}}$-dependent terms, which are interpreted as stochastic fluctuations \cite{Huang:2023eyz}.  To translate \eqref{eq:MSR} to the formalism of \cite{Akyuz:2023lsm,Huang:2023eyz}, one sets $n_{\mathrm{a}}\rightarrow n$ (the density of the U(1) charge), and $\phi_{\mathrm{a}} \rightarrow -\pi$, where $\pi$ is the MSR noise field conjugate to $n$.

The more popular choice in the recent literature, however, has been to integrate out $n_{\mathrm{a}}$ and $n_{\mathrm{r}}$ instead \cite{Crossley:2015evo}.  This choice becomes particularly elegant when we assume that $\hat\mu \propto \mu$, with the former defined in \eqref{eq:dissH} governing the steady state to which the dynamics approaches, as is the case for systems in a thermal steady state $\rho \propto \mathrm{e}^{-\beta H}$, where the same (single-copy Hilbert space) Hamiltonian $H$ generates both time evolution and controls the steady state.  In this setting, we indeed would see from \eqref{eq:HLHR} and $\Phi=\beta H$ that \begin{equation}
    \mu = \beta  \hat\mu. \label{eq:hatmumu}
\end{equation} 
Strictly speaking, systems with conserved energy would also have a diffusing energy density mode, which we are neglecting in our EFT; still, this complication does not change the essential physics we are focusing on here.  Integrating out $n_{\mathrm{a}}$ and $n_{\mathrm{r}}$, we could then write down an EFT subject to the reparameterization symmetries \begin{equation}
\mathcal{L}(\phi_{\mathrm{a}},\phi_{\mathrm{r}}) =     \mathcal{L}(\phi_{\mathrm{a}}+\alpha,\phi_{\mathrm{r}}+f(x)) \label{eq:reparameterization3} 
\end{equation} 
positivity \eqref{eq:positivity}, the modification of \eqref{eq:Hna0} \begin{equation}
    \mathcal{L}(\phi_{\mathrm{a}}=0,\phi_{\mathrm{r}}) = 0,
\end{equation}
and lastly the Kubo-Martin-Schwinger (KMS) symmetry \begin{equation}
    \mathcal{L}(-\phi_{\mathrm{a}}-\mathrm{i}\beta\partial_t \phi_{\mathrm{r}},-\phi_{\mathrm{r}},-t) = \mathcal{L}(\phi_{\mathrm{a}},\phi_{\mathrm{r}},t) + \partial (\cdots) \label{eq:KMS}
\end{equation}
which generalizes \eqref{eq:phiaTRS}, and ensures that up to total derivatives the EFT is time-reversal symmetric.   Notice that the KMS transformation is particularly simple due to \eqref{eq:hatmumu}.  These are precisely the postulates of \cite{Crossley:2015evo} which have led to a modern renaissance in our understanding of the EFT of hydrodynamics.  Their methods led to the following Lagrangian describing diffusion: \begin{equation}
    \mathcal{L} = n\partial_t \phi_{\mathrm{a}} + \mathrm{i}\sigma\nabla \phi_{\mathrm{a}}\cdot \nabla (\phi_{\mathrm{a}}+\mathrm{i}\beta\partial_t\phi_{\mathrm{r}}) + \cdots . \label{eq:SK}
\end{equation}where $\cdots$ denote irrelevant corrections, and the coefficients $n$ and $\sigma$ are functions of $ \mu = \beta \partial_t \phi_{\mathrm{r}}$ subject to the positivity constraint $\sigma\ge 0$. The main result of our paper is an intuitive understanding of the postulates which led to \eqref{eq:SK} -- chiefly the unusual reparameterization symmetry \eqref{eq:reparameterization3} -- which is a clear manifestation of SWSSB, as explained around \eqref{eq:reparameterization2}.  It is easy to see that assuming \eqref{eq:hatmumu}, integrating out the $n$-fields from the EFT \eqref{eq:dissL} and \eqref{eq:dissH} leads to \eqref{eq:SK}; in particular, the quantum/classical field $\phi_{\mathrm{a}}$/$\phi_{\mathrm{r}}$ introduced in the SK formalism arose naturally as canonical conjugates to the strong/weak charge densities $n_{\mathrm{a}}$/$n_{\mathrm{r}}$.

We already discussed various probes and consistency checks for SWSSB coming entirely from equilibrium considerations in Section \ref{sec:swssb}.  The last crucial property of SSB in pure states was the presence of Goldstone's Theorem.
Generalizing the discussion around \eqref{eq:goldstone0}, we now define $A_{\mathrm{a}} = \mathrm{e}^{\mathrm{i}\phi_{\mathrm{a}}}$ and consider 
\begin{align}\label{eq:Grnaphia}
    \int \mathrm{d}^dx \; G^{\mathrm{R}}_{n_{\mathrm{a}}A_{\mathrm{a}}}(x,t)&:=\mathrm{i}\mathrm{\Theta}(t) \int \mathrm{d}^dx (I| [n_{\mathrm{a}}(x,t), A_{\mathrm{a}}(0)]|\rho)  \nonumber\\
    &= \mathrm{i}\mathrm{\Theta}(t) (I| \rho) =\mathrm{i}\mathrm{\Theta}(t)  , 
\end{align} 
where the commutator is evaluated analytically as in \eqref{eq:canonical}, and we are neglecting the compactness of $\phi$ in the hydrodynamic (long-wavelength) limit. 
We remind the reader that $n_{\mathrm{a}}$ and $\phi_{\mathrm{a}}$ are genuine linear operators in the doubled Hilbert space, and that (as stated before) $\phi_{\mathrm{a}}$ is not defined on a single-copy Hilbert space. Similar to the discussion around \eqref{eq:Gr peak}, the imaginary part of the Green's function $G^{\mathrm{R}}_{n_{\mathrm{a}}A_{\mathrm{a}}}(k=0,\omega)$ defined in \eqref{eq:Grnaphia} necessarily has a $\delta(\omega)$ singularity, which signals the presence of gapless modes in the EFT.  These gapless modes are the Nambu-Goldstone bosons of SWSSB, and we now argue that they are precisely the diffusion quasinormal mode \eqref{eq:diffusion} in hydrodynamics.  Taylor expanding $A_{\mathrm{a}}\approx 1 + \mathrm{i}\phi_{\mathrm{a}}+\cdots$, and calculating the Green's function \begin{equation}
          G^{\mathrm{R}}_{n_{\mathrm{a}}A_{\mathrm{a}}}(k,\omega) \approx \mathrm{i} G^{\mathrm{R}}_{n_{\mathrm{a}}\phi_{\mathrm{a}}}(k,\omega) = \frac{-1}{\omega + \ii D k^2},
\end{equation}
 we see that the spectral weight \begin{equation}
  \mathrm{Im} G^{\mathrm{R}}_{n_{\mathrm{a}}A_{\mathrm{a}}}(k,\omega) = \frac{1}{2\chi} G^{\mathrm{S}}_{nn}(k,\omega) = \frac{Dk^2}{\omega^2+(Dk^2)^2}. \label{eq:GRtoGS}
\end{equation}
where $G^{\mathrm{S}}_{nn}(x,t) = \langle \lbrace n(x,t),n(0)\rbrace\rangle $ is the symmetric Green's function in the single-copy Hilbert space, whose value in the hydrodynamic limit is provided in e.g. \cite{Crossley:2015evo}.

The singularity in \eqref{eq:GRtoGS}, \begin{equation}
    G^{\mathrm{S}}_{nn}(k=0,\omega) = 2\pi \chi \times \delta(\omega) ,
\end{equation}
precisely matches the singularity in the Fourier transform of \eqref{eq:Grnaphia}, as described in \eqref{eq:Gr peak}.  A physical understanding of this result follows from \eqref{eq:phiaapprox}, which implies that 
\begin{equation}
    (I| [n_{\mathrm{a}}(x,t), \mathrm{e}^{\mathrm{i}\phi_{\mathrm{a}}(0)}]|\rho)\approx \frac{1}{2\chi} G^{\mathrm{S}}_{nn}(x,t).
\end{equation}
Therefore, so long as $\chi>0$ and we have SWSSB, the Goldstone boson is simply the diffusion mode, which can (and has, routinely) been detected in experiment. The spectral weight in \eqref{eq:Grnaphia} as $\omega \rightarrow 0$ ensures that the diffusion mode has arbitrarily long lifetimes in the thermodynamic limit (on sufficiently long length scales).  

Lastly, we note that \eqref{eq:GRtoGS} is a direct consequence of time-reversal (KMS) symmetry, which further implies a fluctuation-dissipation theorem \cite{Crossley:2015evo,Huang:2023eyz}.

\section{Breaking the weak symmetry}\label{sec:sf}
We can also use our formalism to transparently understand how to build the EFT for a dissipative superfluid.  Such a prescription is well-known in the literature \cite{Crossley:2015evo}, so we will be brief; our point here is just to stress that we can now transparently understand the prescription we summarize below.  To go from a normal fluid to a superfluid, we should clearly relax the reparameterization symmetry \eqref{eq:reparameterization2} down to \begin{equation}
        \mathcal{H}(n_{\mathrm{a}}, n_{\mathrm{r}}, \phi_{\mathrm{a}}, \phi_{\mathrm{r}}) = \mathcal{H}(n_{\mathrm{a}}, n_{\mathrm{r}}, \phi_{\mathrm{a}} + \alpha, \phi_{\mathrm{r}} + \alpha^\prime)
\end{equation}
where $\alpha,\alpha^\prime$ are constants, since now both the strong $\mathrm{U}(1)_{\mathrm{a}}$ and weak $\mathrm{U}(1)_{\mathrm{r}}$ undergo SSB.  We still require \eqref{eq:Hna0}, such that each term in $\mathcal{H}$ requires either a copy of $\nabla \phi_{\mathrm{a}}$ or $n_{\mathrm{r}}$. 

The most relevant terms in $\mathcal{H}$ are \begin{align}
         \mathcal{H} &=  \hat \mu(n_{\mathrm{a}})n_{\mathrm{r}} + K(n_{\mathrm{a}}) \nabla \phi_{\mathrm{r}} \cdot \nabla\phi_{\mathrm{a}} \nonumber\\
         &\quad \quad - \mathrm{i}\sigma(n_{\mathrm{a}}) \nabla \phi_{\mathrm{a}} \cdot \nabla (\phi_{\mathrm{a}}+\mathrm{i}\mu(n_{\mathrm{a}})) + \cdots. \label{eq:dissSF H}
 \end{align}
 After integrating out $n_{\mathrm{r}}$ with the constraint \eqref{eq:dtphir mu}, we arrive at the Lagrangian
 \begin{align}
    \mathcal{L} &= n_{\mathrm{a}}\partial_t \phi_{\mathrm{a}} - K(n_{\mathrm{a}})\nabla \phi_{\mathrm{r}} \cdot \nabla\phi_{\mathrm{a}} \notag \\
    &+ \mathrm{i}\sigma(n_{\mathrm{a}}) \nabla \phi_{\mathrm{a}}\cdot \nabla (\phi_{\mathrm{a}}+\mathrm{i}\mu(n_{\mathrm{a}})). \label{eq:MSR SF}
\end{align}
Varying with respect to $\phi_{\mathrm{a}}$, we obtain the dissipative equation for the superfluid: 
\begin{equation}\label{eq:sf eom}
    \partial_t^2 \phi_{\mathrm{r}} - v_{\mathrm{s}}^2 \nabla^2 \phi_{\mathrm{r}} = \nabla \cdot (D\nabla \p_t\phi_{\mathrm{r}}) + \cdots,
\end{equation}
where we used \eqref{eq:dtphir mu} to write $n_{\mathrm{a}} = \hat\chi \p_t \phi_{\mathrm{r}}$ with $\hat\chi\equiv \p n_a/\p \hat \mu$, and $v_{\mathrm{s}}= \sqrt{K/\hat \chi} $ is the superfluid velocity. The normal modes corresponding to \eqref{eq:sf eom} are given by
\begin{align}
    \omega = \pm v_s k - \ii \frac{D}{2}k^2 + \cdots,
\end{align}
which features a pair of superfluid ``second sound" modes -- the Nambu-Goldstone bosons for the SSB $\mathrm{U}(1)$ in the single-copy Hilbert space -- damped by diffusion.

\section{Outlook}
In this paper, we studied a minimal model of SWSSB for a continuous symmetry -- the emergence of hydrodynamic diffusion in a mixed state.  We showed that the hydrodynamic EFT of diffusion can be derived as a kind of ``superfluid" EFT for SWSSB, which provides a transparent understanding of the origins of all symmetries of hydrodynamic EFTs \cite{Crossley:2015evo}.  Moreover, the gapless hydrodynamic quasinormal modes were the Nambu-Goldstone bosons of SWSSB, which provides a profound new understanding of why hydrodynamic modes must always be long-lived.  Our work also therefore demonstrates the physical insight gained from using new frameworks, such as SWSSB, to understand old problems in many-body physics.  Although we have focused on a minimal model, we do not see any reason why our conclusions do not generalize to more complicated hydrodynamic theories.

One interesting thing that arose in our analysis is that it appeared possible to write down a hydrodynamic EFT for a quantum system which was not in thermal equilibrium.  To do this, one simply does not require \eqref{eq:hatmumu}.  One can still integrate out $n_{\mathrm{a}}$ and $n_{\mathrm{r}}$ to obtain a hydrodynamic EFT, but the KMS transformation \eqref{eq:KMS} will be, in general, more complicated: \begin{equation}
    \phi_{\mathrm{a}} \rightarrow -\phi_{\mathrm{a}}- \mathrm{i} \mu \left(\hat\mu^{-1}\left(\partial_t \phi_{\mathrm{r}}\right)\right) + \cdots.
\end{equation}
This suggests that following \cite{Huang:2023eyz}, one can systematically build field theories for nonthermal quantum many-body systems, such as random unitary circuits \cite{Ogunnaike:2023qyh,khemani,tibor} where hydrodynamic phenomena have been extensively studied.  It is an interesting open question whether such field  theories can capture any of the nonclassical phenomena, such as dynamics with measurement and feedback \cite{NahumMeasInduced,FisherQuantumZeno,ChoiQEC,Ludwig2020,RoySteering,lee2022decoding,FriedmanMIPT}, described in the systematic framework of \cite{Guo:2024ycr}.

 \emph{Note added.}--- As we were completing this manuscript, another manuscript \cite{Gu:2024wgc}, which also includes discussions of diffusion and SWSSB, appeared.

\section*{Acknowledgements}
We thank Drew Potter, Chong Wang, Cenke Xu, and Yichen Xu for useful discussions.

This work was supported by the Alfred P. Sloan Foundation under Grant FG-2020-13795 (AL), the National Science Foundation under CAREER Grant DMR-2145544 (AL), the Department of Energy under Basic Energy Sciences Award DE-SC0014415 (MQ) and Quantum Pathfinder Award DE-SC0024324 (JHZ), and the Heising-Simons Foundation under Grant 2024-4848 (XH, AL).  XH and JHZ thank the Kavli Institute for Theoretical Physics for hospitality; KITP is supported in part by the Heising-Simons Foundation, the Simons Foundation, and the National Science Foundation under Grant PHY-2309135.

\appendix

\section{1+1d gapless Dirac fermion}\label{app:dirac}

A symmetric quantum system can have nontrivial excitations if the symmetry is anomalous. For example, the Dirac fermion in 1+1 dimensions could be gapless without breaking any symmetry. This example does not appear to fall into our categories as it has neither SSB nor reparametrization symmetry. However, we show here that the EFT \emph{can} be written in a way with manifest reparametrization symmetry. The key insight is that the Dirac fermion has \emph{two} $U(1)$ symmetries with a mixed anomaly; treating the two $U(1)$ symmetries on equal footing leads to an action with manifest reparametrization symmetry.
In the following, we will work with the $1+1d$ compact boson, which is related to the $1+1$d Dirac fermion by bosonization, and is completely equivalent at the level of hydrodynamics. 

We begin with the action for the compact boson with the identification $\varphi \sim \varphi + 2\pi$
\begin{equation} \label{eq:boson S}
    S[\varphi] = \frac{R^2}{4\pi} \int \ud x \ud t~ (\p_t \varphi)^2 - (\p_x \varphi)^2.
\end{equation}
The radius $R = \sqrt{2}$ corresponds to the bosonized free Dirac fermion. This action is manifestly invariant under the symmetry $\varphi \to \varphi + c$. It will be helpful to review how T-duality rewrites the action in terms of a dual field $\theta$ which is manifestly invariant under the symmetry $\theta \to \theta + c$. Introduce fields $j_t = \p_t \varphi$ and $j_x = \p_x \varphi$. As the dependence of $S[\varphi]$ on $\varphi$ only enters through $j$, we would like to replace the path integral over $\varphi$ with a path integral over $j$; however, in doing so we need to restrict to field configurations satisfying $\p_t j_x = \p_x j_t$. This is accomplished with a Lagrange multiplier $\theta$, leading to the action 
\begin{equation} \label{eq:jtheta S}
    S[j, \theta] = \int \ud x \ud t~ \frac{R^2}{4\pi} (j_t^2 - j_x^2) + \frac{\theta}{2\pi} (\p_t j_x - \p_x j_t). 
\end{equation}
Compactness of $\theta \sim \theta + 2\pi$ ensures that vortex configurations of $j$ are included in the path integral. Integrating out $j$ gives the action in terms of the dual $\theta$
\begin{equation}
    S[\theta] = \frac{1}{4\pi R^2} \int \ud x \ud t~ (\p_t \theta)^2 - (\p_x \theta)^2
\end{equation}
where the symmetry $\theta \to \theta + c$ is manifest. This action, like \eqref{eq:boson S}, does not have a reparametrization invariance. 

Let us return to \eqref{eq:jtheta S} and consider the modified action
\begin{equation}
    S[\varphi, j_x, \theta] = \int \ud x \ud t~ \frac{R^2}{4\pi} ((\p_t \varphi)^2 - j_x^2) + \frac{\theta}{2\pi} (\p_t j_x - \p_x \p_t \varphi) 
\end{equation}
where we have replaced $j_t$ with $\p_t \theta$ but kept $j_x$ in anticipation of integrating it out. Interestingly, this action has two reparametrization symmetries 
\begin{equation}
    \varphi \to \varphi + f(x), ~ \theta \to \theta + g(x)
\end{equation}
for time-independent functions $f(x)$ and $g(x)$. Integrating out $j_x$, we obtain 
\begin{equation} \label{eq:phi theta S}
    S[\varphi, \theta] = \int \ud x \ud t~ \frac{1}{2\pi} \p_t \varphi \p_x \theta + \frac{R^2}{4\pi} (\p_t \varphi)^2 + \frac{1}{4\pi R^2} (\p_t \theta)^2.
\end{equation}
This can be thought of implementing ``half" of T-duality to rewrite the action in terms of both $\varphi$ and $\theta$ variables in such a way that both $\varphi$ and $\theta$ have a shift symmetry. This action is in standard ``hydrodynamic form": the first term characterizes the mixed anomaly between the two $U(1)$ symmetries, and the remaining terms depend only on time derivatives of the phase fields, which is standard when the symmetries are unbroken. This occurred naturally when both $U(1)$ symmetries were made manifest in the action. Note that the reparametrization symmetry is not accidental: relaxing the symmetry and allowing terms such as $(\p_x \theta)^2$ changes the action and makes the theory different from the compact boson and Dirac fermion.

Some care is required in going between this action and the original compact boson action that we started with. Varying with respect to $\theta$ of \eqref{eq:phi theta S} gives
\begin{align}
    \p_t \left(\p_x \varphi + \frac{1}{R^2} \p_t \theta \right) = 0
\end{align}
which is solved by setting $\p_x \varphi + \frac{1}{R^2} \p_t \theta = F(x)$. Using the equations of motion to integrate out $\theta$ and shifting $S\to S - \int\ud x\ud t~ \frac{1}{2\pi}F(x)\p_t\theta$ by a total derivative term gives
\begin{equation}
    S[\varphi] = \frac{R^2}{4\pi} \int \ud x \ud t~ (\p_t \varphi)^2 - (\p_x \varphi - F)^2 
\end{equation}
which reduces to the compact boson action when $F = 0$. Now, $S[\varphi]$ can be invariant under the reparameterization symmetry $\varphi\to \varphi + f(x) $ if we also shift $F \to F - \p_x f(x)$. This suggests that different initial spatial profiles of $\varphi$ do not change the physics of Dirac fermions; indeed, one can add a constant phase to the textbook bosonization dictionary \cite{Shankar_2017} without causing any change to the correlation function and the commutation relation. By contrast, this initial condition will cost energy in superfluids where the reparameterization symmetry is absent.

\bibliography{SWSSB}

\end{document}